\documentclass[11pt]{JHEP3}
\usepackage{graphicx}
\usepackage{cite,./mcite}
\usepackage{url,relsize}
\usepackage{xspace,amsmath,amssymb}
\usepackage{snapshot}

\IfFileExists{microtype.sty}{%
  \RequirePackage{microtype}%
}{}

\newcommand{\p}{\ensuremath{\vec{p}}\xspace}
\newcommand{\pzero}{\ensuremath{\vec{p}_0}\xspace}
\newcommand{\pprime}{\ensuremath{\vec{p}^{\,\prime}}\xspace}
\newcommand{\Nmin}[1]{\ensuremath{N_\text{min}^{\,(#1)}}\xspace}

\newcommand{\pT}{\ensuremath{p_\perp}\xspace}
\newcommand{\LambdaQCD}{\ensuremath{\Lambda_\text{QCD}}\xspace}
\newcommand{\ofOrder}[1]{\ensuremath{\mathcal{O}(#1)}\xspace}

\newcommand{\prog}[1]{{\smaller\textsf{#1}}}
\let\myurl\url
\renewcommand{\url}[1]{{\smaller\textsf{#1}}}

\renewcommand{\url}[1]{{\smaller{\myurl{#1}}}}

\renewcommand{\prog}[1]{\textsf{#1}}

\title{Tools for event generator tuning and validation}
\author{Andy Buckley\\Institute for Particle Physics Phenomenology,\\ Durham University, UK}
\abstract{
  I describe the current status of MCnet tools for validating the performance of
  event generator simulations against data, and for tuning their
  phenomenological free parameters. For validation, the Rivet toolkit is now a
  mature and complete system, with a large library of prominent benchmark
  analyses. For tuning, the Professor system has recently completed its first
  tunes of Pythia 6, with substantial improvements on the existing default tune
  and potential to greatly aid the setup of new generators for LHC studies.
}

\begin{document}
\maketitle

% \begin{abstract}
%   I describe the current status of MCnet tools for validating the performance of
%   event generator simulations against data, and for tuning their
%   phenomenological free parameters. For validation, the Rivet toolkit is now a
%   mature and complete system, with a large library of prominent benchmark
%   analyses. For tuning, the Professor system has recently completed its first
%   tunes of Pythia 6, with substantial improvements on the existing default tune
%   and potential to greatly aid the setup of new generators for LHC studies.
% \end{abstract}

\section{Introduction}

It is an inevitable consequence of the physics approximations in Monte Carlo
event generators that there will be a number of relatively free parameters which
must be tweaked if the generator is to describe experimental data. Such
parameters may be found in most aspects of generator codes, from choices of
\LambdaQCD and \pT cutoff in the perturbative parton cascade, to the
non-perturbative hadronisation process. These latter account for the majority of
parameters, since the models are deeply phenomenological, typically invoking a
slew of numbers to describe not only the kinematic distribution of \pT in hadron
fragmentation, but also baryon/meson ratios, strangeness and $\{\eta, \eta'\}$
suppression, and distribution of orbital angular
momentum~\cite{pythia,herwig,herwigpp,sherpa}. The result is a proliferation of
parameters --- of which between \ofOrder{\text{10--30}} may be of particular importance
for physics studies.

Apart from rough arguments about their typical scale, these parameters are
freely-floating: they must be matched to experimental data for the generator to
perform well. Additionally, it is important that this tuning is performed
against a wide range of experimental analyses, since otherwise parameters to
which the selective analyses are insensitive will wander freely and may drive
unconsidered observables to bad or even unphysical places. This requires a
systematic and global approach to generator tuning: accordingly, I will
summarise the current state of tools for systematically validating and tuning
event generator parameters, and the first results of such systematic tunings.

\section{Validation tools: Rivet}

The Rivet library is a successor to the successful HERA-oriented generator
analysis library, HZTool~\cite{hztool}. Like its predecessor, the one library
contains both a library of experimental analyses and tools for calculating
physical observables. It is written in object-oriented C++ and there is strong
emphasis on the following features:
\begin{itemize}
\item strict generator-independence: analyses are strictly performed on
  HepMC~\cite{hepmc} event record objects with no knowledge of or ability to
  influence the generator behaviour;
\item experimental reference data files are included for each standard analysis,
  and are used to ensure that analysis data binnings match their experimental
  counterparts as well as for fit comparisons;
\item computational results are automatically cached for use between different
  analyses, using an infrastructure mechanism based on ``projection'' classes;
\item clean, transparent and flexible programming interface: while much of the
  complexity is hidden, analyses retain a clear algorithmic structure rather
  than attempting to hide everything behind ``magic'' configuration files.
\end{itemize}
\goodbreak

\noindent
The ``projection'' objects used to compute complex observables are now a fairly
complete set:

\nobreak
\begin{itemize}
\item various ways to obtain final state particles: all, charged only, excluding
  certain particles, with \pT and rapidity cuts, etc.;
\item event shapes: sphericity, thrust, Parisi C \& D parameters, jet hemispheres;
\item jet algorithms: CDF and D\O{} legacy cones, Durham/\textsc{Jade}, and
  $k_\perp$, anti-$k_\perp$, \textsc{SISCone}, CDF ``\textsc{jetclu}'' etc. from
  FastJet~\cite{fastjet};
\item miscellaneous: jet shapes, isolation calculators, primary and secondary
  vertex finders, DIS kinematics transforms, hadron decay finder, etc.
\end{itemize}

\noindent
The set of standard analyses has also grown with time and is now particularly
well-populated with analyses from the LEP and Tevatron experiments:

\begin{itemize}
\item LEP: \textsc{Aleph} and \textsc{Delphi} event shape analyses;
  \textsc{Aleph}, \textsc{Delphi} and PDG hadron multiplicities, strange
  baryons; \textsc{Delphi} and \textsc{Opal} b-fragmentation analyses;
\item Tevatron: CDF underlying event analyses (from 2001, 2004 \& 2008); CDF and
  D\O{} EW boson \pT analyses; CDF and D\O{} QCD colour coherence, jet
  decorrelation, jet shapes, Z+jets, inclusive jet cross-section;
\item HERA: H1 energy flow and charged particle spectra; ZEUS dijet
  photoproduction.
\end{itemize}

\noindent
In addition, users can write their own analyses using the Rivet projections
without needing to modify the Rivet source, by using Rivet's plugin system. We
encourage such privately-implemented analyses to be submitted for inclusion in
the main Rivet distribution, and would particularly welcome QCD analyses from
HERA, b-factory and RHIC p-p experiments.

While Rivet is primarily a library which can be used from within any analysis
framework (for example, it is integrated into the \textsc{Atlas} experiment's
framework), the primary usage method is via a small executable called
\prog{rivetgun}. This provides a frontend for reading in HepMC events from plain
text dump files and also for running generators ``on the fly'' via the AGILe
interface library. This latter approach is particularly nice because there is no
need to store large HepMC dump files and the corresponding lack of file I/O
speeds up the analysis by a factor $\sim\!\ofOrder{10}$. In this mode, Rivet is
ideal for parameter space scans, since generator parameters can be specified by
name on the \prog{rivetgun} command line and applied without
recompilation. AGILe currently supports API-level interfaces to the Fortran
\textsc{Herwig} 6~\cite{herwig} and Pythia 6~\cite{pythia} generators (combined
with the AlpGen~\cite{alpgen} MLM multi-jet merging generator, the
\textsc{Charybdis} black hole generator~\cite{charybdis}, and the \textsc{Jimmy}
hard underlying event generator~\cite{jimmy} for \textsc{Herwig}), plus the C++
generators Herwig++~\cite{herwigpp}, Sherpa~\cite{sherpa}, and Pythia
8~\cite{pythia8}.

At the time of writing, the current version of Rivet is 1.1.1. The main
framework benefits of the 1.1.x series over 1.0.x are a safer and simpler
mechanism for handling projection objects (massively simplifying many analyses),
better compatibility of the AGILe loader with the standard LCG Genser packaging
and a large number of new and improved analyses and projections. A ``bootstrap''
script is provided for easy setup. Anyone interested in using Rivet for
generator validation should first visit the website
\url{http://projects.hepforge.org/rivet/}.

Rivet is now a stable and powerful framework for generator analysis and we are
looking forward to its increasing r\^ole in constraining generator tunings for
background modelling in LHC high-\pT physics. Future versions will see
improvements aimed at high-statistics validation simulations, such as
histogramming where statistical error merging is automatically correct, as well
as the addition of more validation analyses.

\section{Tuning tools: Professor}

While Rivet provides a framework for comparing a given generator tuning to a
wide range of experimental data, it provides no intrinsic mechanism for
improving the quality of that tune. Historically, the uninspiring task of tuning
generator parameters to data ``by eye'' has been the unhappy lot of experimental
researchers, %
%\footnote{I would say ``graduate students'', but there is at least
% one high-profile exception to that rule.}
with the unsystematic nature of the study reflecting that significant
improvements in quality of both life and tuning would have been possible. %
%\footnote{Another notable exception has been the tuning
%  of Herwig++ 2.2.0, which was mainly based on a massive random grid scan using
%  Grid computing resources. Even with the Grid CPU power, this was only
%  realistically possible because of low batch farm occupancy at the time.} 
This call for an automated and systematic approach to tuning is taken up by a
second new tool: Professor. This is written in Python code as a set of
factorised scripts, using the SciPy numerical library~\cite{scipy} and an
interface to \prog{rivetgun}.

The rough formalism of systematic generator tuning is to define a goodness of
fit function between the generated and reference data, and then to minimise that
function. The intrinsic problem is that the true fit function is certainly not
analytic and any iterative approach to minimisation will be doomed by the
expense of evaluating the fit function at a new parameter-space point: this may
well involve ten or more runs of the generator with 200k--2M events per
run. Even assuming that such runs can be parallelised to the extent that only
the longest determines the critical path, an intrinsically serial minimisation
of \ofOrder{1000} steps will still take many months. This is clearly not a
realistic strategy!

The Professor approach, which is the latest in a lengthy but vague history of
such efforts\cite{delphi-prof,delphi-prof2}, is to parameterise the fit function with a
polynomial. In fact, since the fit function itself is expected to be complex and
not readily parameterisable, there is a layer of indirection: the polynomial is
actually fitted to the generator response of each observable bin, $\text{MC}_b$
to the changes in the $n$-element parameter vector, \p. To account for
lowest-order parameter correlations, a second-order polynomial is used,
\begin{align}
  \label{eq:poly}
  \text{MC}_b(\p\,) 
  \approx f^{(b)}(\p\,)
  = \alpha^{(b)}_0 + \sum_i \beta^{(b)}_i \, p^\prime_i 
  + \sum_{i \le j} \gamma^{(b)}_{ij} \, p^\prime_i \, p^\prime_j
  ,
\end{align}
where the shifted parameter vector $\pprime \equiv \p - \pzero$, with \pzero
chosen as the centre of the parameter hypercube. A nice feature of using a
polynomial fit function, other than its general-purpose robustness, is that the
actual choice of the \pzero is irrelevant: the result of a shift in central
value is simply to redefine the coefficients, rather than change the functional
form, but choosing a central value is numerically sensible.

% \begin{figure}[t]
%   \centering
%   \includegraphics[width=0.2\textwidth]{sample-hcube}
%   \qquad
%   \includegraphics[width=0.3\textwidth]{pythia-thrust-chi2-scan}
%   \caption{Random sampling}
%   \label{fig:sampling}
% \end{figure}

The coefficients are determined by randomly sampling the generator from $N$
parameter space points in an $n$-dimensional parameter hypercube defined by the
user. Each sampled point may actually consist of many generator runs, which are
then merged into a single collection of simulation histograms. A simultaneous
equations solution is possible if the number of runs is the same as the number
of coefficients between the $n$ parameters, i.e. $N = \Nmin{n} = (2 + 3n +
n^{2})/2$. However, using this minimum number of runs introduces a systematic
uncertainty, as we certainly do not expect the bin MC response to be a perfect
polynomial. Here we are helped by the existence of the Moore--Penrose
pseudoinverse: a generalisation of the normal matrix inverse to non-square
matrices with the desirable feature that an over-constrained matrix will be
inverted in a way which gives a least-squares best fit to the target
vector. Even more helpful is that a standard singular value decomposition (SVD)
procedure can be used to deterministically implement the pseudoinverse
computation. Hence, we phrase the mapping on a bin-by-bin basis from
coefficients $C$ to generator values $V$ as $P C = V$, where $P$ is the
parameter matrix to be pseudo-inverted. For a two parameter case, parameters
$\in \{x,y\}$, the above may be explicitly written as
\newcommand{\columnfill}{%
  \begin{pmatrix}
    \alpha_0 \\ \beta_x \\ \beta_y \\ \gamma_{xx} \\ \gamma_{xy} \\ \gamma_{yy}
  \end{pmatrix}
}
\begin{align}
  \underbrace{
    \vphantom{\columnfill}
    \begin{pmatrix}
      1 & x_1 & y_1 & x^2_1 & x_1y_1 & y^2_1 \\
      1 & x_2 & y_2 & x^2_2 & x_2y_2 & y^2_2 \\
      &     &     &  \vdots   &   &
    \end{pmatrix}
  }_{P\text{ (sampled param sets)}}
  \underbrace{
    \begin{pmatrix}
      \alpha_0 \\ \beta_x \\ \beta_y \\ \gamma_{xx} \\ \gamma_{xy} \\ \gamma_{yy}
    \end{pmatrix}
  }_{C\text{ (coeffs)}}
  = 
  \underbrace{
    \vphantom{\columnfill}
    \begin{pmatrix}
      v_1 \\ v_2 \\ \vdots
    \end{pmatrix}
  }_{V\text{ (values)}}
\end{align}
where the numerical subscripts indicate the $N$ generator runs. Note that the
columns of $P$ include all $\Nmin{2} = 6$ combinations of parameters in the
polynomial, and that $P$ is square (i.e. minimally pseudo-invertible) when $N =
\Nmin{n}$.  Then $C = \tilde{\mathcal{I}}[P] \, V$, where $\tilde{\mathcal{I}}$
is the pseudoinverse operator.

Now that we have, in principle, a good parameterisation of the generator
response to the parameters, $\p$, for each observable bin, $b$, it remains to
construct a goodness of fit (GoF) function and minimise it. We choose the
$\chi^2$ function, but other GoF measures can certainly be used. Since the
relative importance of various distributions in the observable set is a
subjective thing --- given 20 event shape distributions and one charged
multiplicity, it is certainly sensible to weight up the multiplicity by a factor
of at least 10 or so to maintain its relevance to the GoF measure --- we include
weights, $w_{\mathcal{O}}$, for each observable, $\mathcal{O}$, in our $\chi^2$
definition:
\begin{align}
  \chi^2(\p\,) = 
  \sum_{\mathcal{O}} w_{\mathcal{O}} \sum_{b \, \in \, \mathcal{O}} 
  \frac{ (f_b(\p\,) - \mathcal{R}_b)^2 }{ \Delta^2_b },
\end{align}
where $\mathcal{R}_b$ is the reference value for bin $b$ and the total error
$\Delta_b$ is the sum in quadrature of the reference error and the statistical
generator errors for bin $b$ --- in practise we attempt to generate enough data
that the MC error is much smaller than the reference error for all bins.

The final stage of our procedure is to minimise this parameterised $\chi^2$
function. It is tempting to think that there is scope for an analytic global
minimisation at this order of polynomial, but not enough Hessian matrix elements
may be calculated to constrain all the parameters and hence we must finally
resort to a numerical minimisation. We have implemented this in terms of
minimisers from SciPy and also PyMinuit~\cite{pyminuit}, with the latter's
initial parameter space grid scan making it our preferred choice.

\begin{figure}[t]
  \centering
  \includegraphics[width=0.4\textwidth]{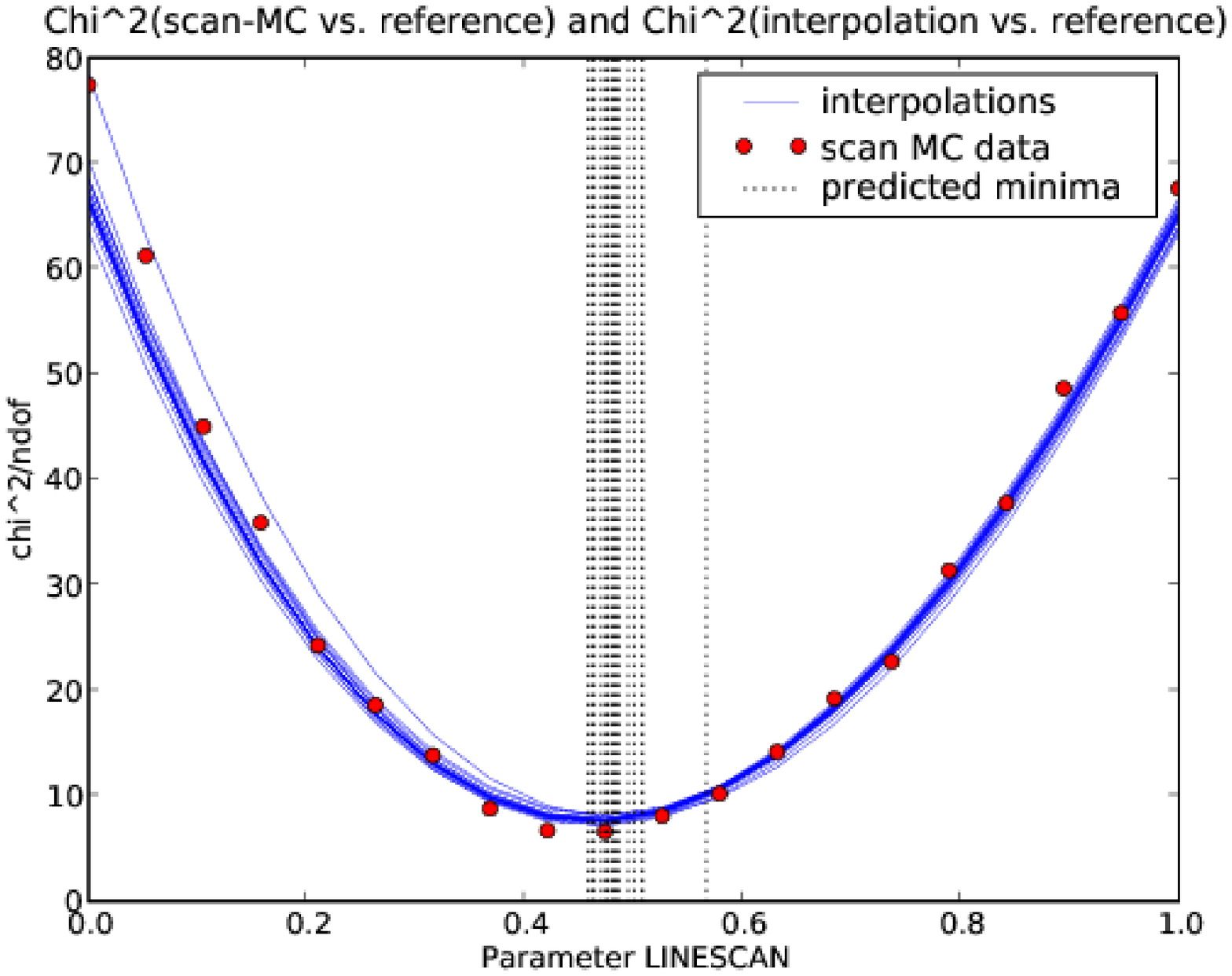}
  \qquad
  \includegraphics[width=0.4\textwidth]{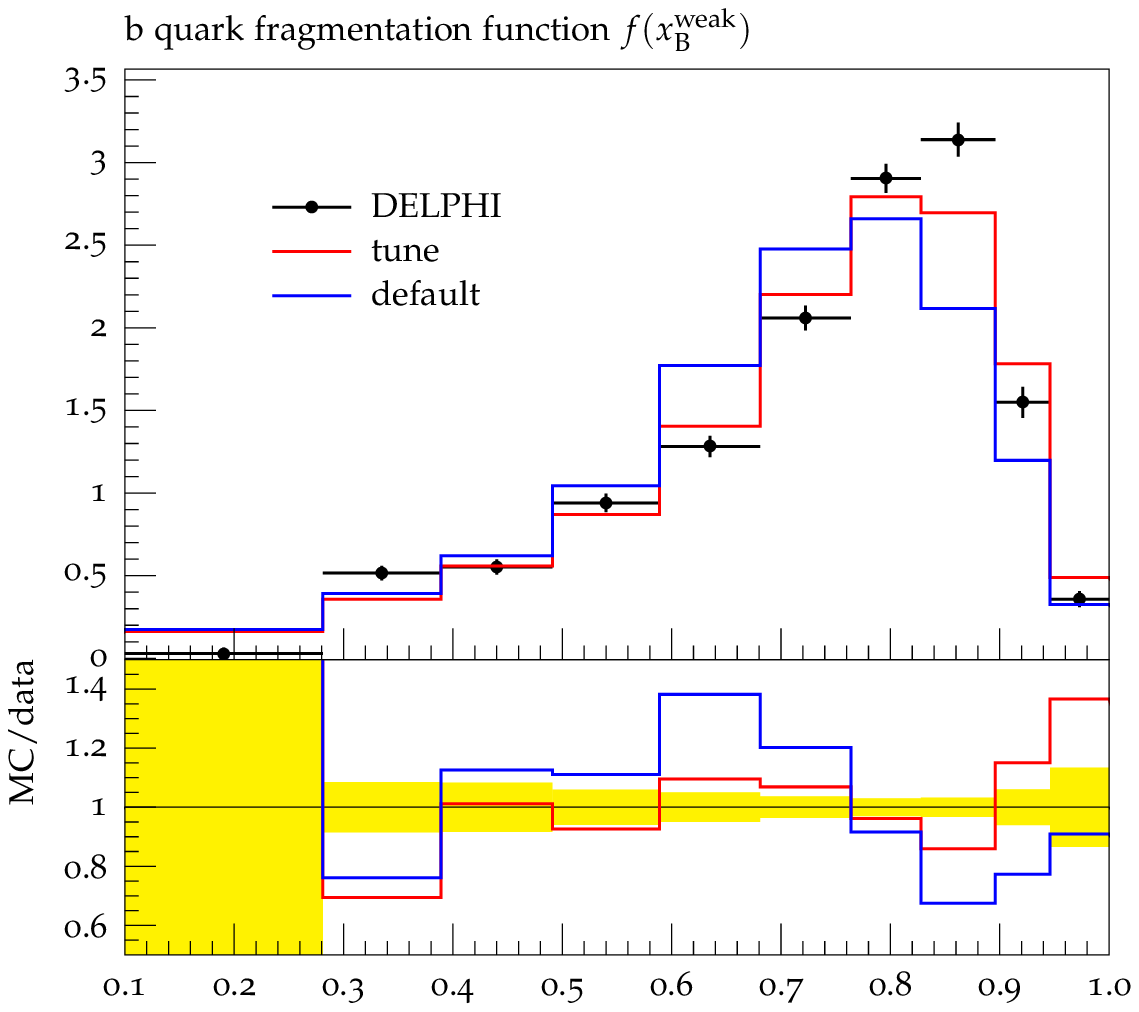}
  %\qquad
  %\includegraphics[width=0.2\textwidth]{d04-x01-y01}
  %\qquad
  %\includegraphics[width=0.4\textwidth]{d05-x01-y01}
  \caption{(a) Parameter space line scan in $\chi^2$, showing the agreement
    between Professor's predicted values (blue lines) and the true values (red
    dots). (b) Pythia 6 b-fragmentation functions, showing the improvements
    obtained using Professor (red) to tune the Bowler parameterisation against
    the default (blue).}
  \label{fig:prof}
\end{figure}

% \begin{figure}[t]
%   \centering
%    \includegraphics[width=0.2\textwidth]{scan-hcube}
%    \qquad
%   \includegraphics[width=0.4\textwidth]{prof-linescan-delphi3d}
%   \caption{Scans}
%   \label{fig:scans}
% \end{figure}

Finally, on obtaining a predicted best tune point from Professor, it is prudent
to check the result. This can be done directly with \prog{rivetgun}, and
Professor also has a line scan feature which allows scans along arbitrary
straight lines in parameter space, which is useful to verify that the $\chi^2$
behaves as interpolated and to explicitly compare default tunes to predicted
tunes. Such a line scan can be seen in Fig. \ref{fig:prof}(a). We have explicitly
checked the robustness of the polynomial and the random distribution of sampling
points against various skewed test distributions and the behaviour is robust. We
have also found it to be useful to over-sample by a considerable fraction, and
then to perform the $\chi^2$ minimisation for a large number of distinct
run-combinations, $\Nmin{n} < N_{\text{tune}}\le N$, which gives a systematic
control on interpolation errors and usually a better performance than just using
$N_{\text{tune}} = N$.\footnote{Note that the tuning runs need a significant
  degree of variation, i.e. $N_{\text{tune}} \ll N$ for most of the tune
  run-combinations.}

% \begin{figure}[t]
%   \centering
%   %\includegraphics[width=0.2\textwidth]{d01-x01-y01}
%   %\quad
%   \includegraphics[width=0.4\textwidth]{d02-x01-y01}
%   %\quad
%   %\includegraphics[width=0.2\textwidth]{d04-x01-y01}
%   \qquad
%   \includegraphics[width=0.4\textwidth]{d05-x01-y01}
%   \caption{b-fragmentation functions}
%   \label{fig:py6plots}
% \end{figure}

The focus in testing and commissioning the Professor system has until recently
been focused on Pythia 6 tunes against LEP data\cite{prof:py6}. Here we were
able to interpolate and minimise up to 10 parameters at a time for roughly 100
distributions, but beyond this the minimisation time became large and we were
less happy with the minima. Eventually we decided to split the tuning into a
two-stage procedure where flavour-sensitive fragmentation parameters were tuned
first to provide a base on which to tune the semi-factorised kinematic
parameters of the shower and hadronisation. The result has been a dramatic
improvement of the Pythia 6 identified particle multiplicity spectra, without
losing the event shape descriptions (originally tuned by \textsc{Delphi}'s
version of the same procedure), and a major improvement of the b-fragmentation
function as seen in Fig. \ref{fig:prof}(b).\footnote{Note that interpolation
  methods cannot deal with discrete settings such as the choice of functional
  form of b-fragmentation function. This required several parallel tunes with
  different values of the discrete parameter.} This tune will be adopted as the
default parameter set for the next release of Pythia 6.

\section{Conclusions}

To conclude, the situation is looking positive for MC generator tuning at
present: the Rivet and Professor tools are now in a state where they can be used
to achieve real physics goals and the Pythia 6 tune described here (using both
tools) has been a significant success. Development plans in the near future are
very much aimed at getting the same tuning machinery to work for hadron collider
studies, in particular initial state radiation (ISR) and underlying event (UE)
physics. We aim to present tunes of C++ generators to LEP data shortly, along
with first studies of interpolation-based tunes to CDF underlying event
data. Finally, we are keen to constrain fragmentation and UE hadron physics for
the LHC, using b-factory, RHIC and early LHC data.

%------------------------------------------------------------------------------
%       Bibliography
%------------------------------------------------------------------------------
%\bibliographystyle{elsart-num}
\bibliographystyle{h-physrev3}
{\raggedright
\bibliography{heralhc}

\begin{thebibliography}{10}

\bibitem{pythia}
{T.~Sjostrand and S.~Mrenna and P.~Skands},
\newblock JHEP {\bf 05}, 026 (2006), hep-ph/0603175.
%%CITATION = HEP-PH/0603175;%%

\bibitem{herwig}
{B.~Webber and others},
\newblock JHEP {\bf 01}, 010 (2001), hep-ph/0011363,
\newblock HERWIG Collaboration.
%%CITATION = HEP-PH/0011363;%%

\bibitem{herwigpp}
P.~Richardson {\em et~al.},
\newblock (2008), 0803.0883,
\newblock Herwig++ Collaboration.
%%CITATION = 0803.0883;%%

\bibitem{sherpa}
{F.~Krauss and others},
\newblock JHEP {\bf 02}, 056 (2004), hep-ph/0311263,
\newblock Sherpa Collaboration.
%%CITATION = HEP-PH/0311263;%%

\bibitem{hztool}
J.~Bromley {\em et~al.},
\newblock (1995),
\newblock ZEUS and H1 Collaborations.

\bibitem{hepmc}
{M.~Dobbs and J.~B.~Hansen},
\newblock Comput. Phys. Commun. {\bf 134}, 41 (2001).
%%CITATION = CPHCB,134,41;%%

\bibitem{fastjet}
{M.~Cacciari and G.~Salam and G.Soyez},
\newblock (2006), hep-ph/0607071.
%%CITATION = HEP-PH/0607071;%%

\bibitem{alpgen}
M.~Mangano {\em et~al.},
\newblock JHEP {\bf 07}, 001 (2003), hep-ph/0206293.
%%CITATION = HEP-PH/0206293;%%

\bibitem{charybdis}
C.~Harris, P.~Richardson, and B.~Webber,
\newblock JHEP {\bf 08}, 033 (2003), hep-ph/0307305.
%%CITATION = HEP-PH/0307305;%%

\bibitem{jimmy}
J.~Butterworth, J.~Forshaw, and M.~Seymour,
\newblock Z. Phys. {\bf C72}, 637 (1996), hep-ph/9601371.
%%CITATION = HEP-PH/9601371;%%

\bibitem{pythia8}
T.~Sjostrand, S.~Mrenna, and P.~Skands,
\newblock Comput. Phys. Commun. {\bf 178}, 852 (2008), 0710.3820.
%%CITATION = 0710.3820;%%

\bibitem{scipy}
SciPy website: {http://www.scipy.org}.

\bibitem{delphi-prof}
DELPHI~Collaboration, K.~Hamacher {\em et~al.},
\newblock Z. Phys. {\bf C73}, 11 (1996).
%%CITATION = ZEPYA,C73,11;%%

\bibitem{delphi-prof2}
{K.~Hamacher and M.~Weierstall},
\newblock (1995), hep-ex/9511011.
%%CITATION = HEP-EX/9511011;%%

\bibitem{pyminuit}
{PyMinuit} website: {http://code.google.com/p/pyminuit/}.

\bibitem{prof:py6}
A.~Buckley {\em et~al.},
\newblock Professor Collaboration, in preparation.

\end{thebibliography}
}
\end{document}